\documentclass[11pt,a4paper]{article}
\usepackage{graphicx}
\usepackage{jcappub}

\title{Constraining the neutrino magnetic dipole moment from
white dwarf pulsations}

\author[a,b]{A. H. C\'orsico,}
\author[a,b]{L. G. Althaus,}
\author[b,c]{M. M. Miller Bertolami,}
\author[d]{S. O. Kepler,}
\author[e,f]{E. Garc\'ia-Berro}

\affiliation[a]{Grupo de Evoluci\'on Estelar y Pulsaciones. 
                Facultad de Ciencias Astron\'omicas y Geof\'{\i}sicas,
                Universidad  Nacional de La Plata,
                Paseo del  Bosque s/n,
               (1900) La Plata,
                Argentina}
\affiliation[b]{Instituto de Astrof\'{\i}sica La Plata,
                CONICET-UNLP,
                Argentina}
\affiliation[c]{Max-Planck-Institut f\"ur Astrophysik,
                Karl-Schwarzschild-Str. 1, 8574, Garching
                Germany}
\affiliation[d]{Departamento de Astronomia,
                Universidade Federal do Rio Grande do Sul,
                Av. Bento Goncalves 9500,
                Porto Alegre 91501-970, RS,
                Brazil}
\affiliation[e]{Departament de F\'isica Aplicada, 
                Universitat Polit\`ecnica de Catalunya, 
                c/Esteve Terrades 5, 08860, 
                Castelldefels, Spain}
\affiliation[f]{Institute for Space Studies of Catalonia, c/Gran 
                Capit\`a 2-4, Edif. Nexus 104, 08034, 
                Barcelona, Spain}

\emailAdd{acorsico@fcaglp.unlp.edu.ar}

\abstract{Pulsating white dwarf stars can be used  as astrophysical
  laboratories to constrain the properties of  weakly  interacting
  particles.  Comparing  the  cooling  rates  of these stars with  the
  expected   values from  theoretical models  allows us  to search for
  additional   sources of  cooling due to the  emission of axions,
  neutralinos,  or neutrinos with magnetic dipole moment.  
  In this  work,  we derive an upper bound to
  the neutrino magnetic  dipole moment ($\mu_\nu$) using an  estimate
  of the rate of period  change of the pulsating DB white  dwarf star
  PG 1351+489. We employ  state-of-the-art evolutionary and pulsational
  codes  which allow us to perform   a  detailed   asteroseismological
  period fit    based   on  fully DB white dwarf evolutionary
  sequences. Plasmon neutrino emission is the dominant cooling
  mechanism for this class of hot pulsating white  dwarfs, and so it is
  the main contributor to the  rate of change  of period   with time
  ($\dot{\Pi}$) for  the DBV class. Thus, the inclusion  of an anomalous
  neutrino emission through a non-vanishing magnetic dipole moment
  in  these sequences notably  influences  the
  evolutionary   timescales,  and  also  the expected pulsational
  properties of the DBV stars.   By comparing  the
  theoretical  $\dot{\Pi}$  value  with  the rate of change of 
  period  with time of PG 1351+489, we  assess  the
  possible existence  of additional cooling by neutrinos with magnetic 
  dipole moment.  Our
  models suggest the  existence of some additional cooling  in this
  pulsating  DB white dwarf,  consistent with   a non-zero magnetic
  dipole moment with an upper limit of $\mu_\nu
  \lesssim 10^{-11} \mu_{\rm B}$. This bound  is somewhat
  less restrictive than, but still compatible with, other   limits
  inferred from  the white dwarf luminosity function  or from the
  color-magnitude diagram  of the Globular cluster M5. 
  Further improvements of the measurement of the rate of period change  
  of the dominant pulsation mode of PG 1351+489 will be necessary to  
  confirm our bound.} 

\notoc

  
\begin{document}

\maketitle  


\section{Introduction}
\label{introduction}

The existence of neutrinos was first postulated by W. Pauli in  1930
to explain the  conservation of energy, momentum, and angular momentum
in $\beta$-decays \cite{1996slfp.book.....R}.   According to the
Standard Model  of particle physics, neutrinos are massless,
electrically neutral, have zero dipole moment, and zero decay rate.
In order to explain the observed neutrino mixing beyond the  Standard
Model, it was necessary to extend this simple picture, allowing for
non-vanishing neutrino masses and mixings, electromagnetic couplings,
neutrino decays and other effects \cite{1996slfp.book.....R}.

Neutrino emission takes place in a number of astrophysical contexts
and constitutes an energy-loss channel in a variety of stellar
configurations,  from  low-mass red giants and horizontal branch
stars to white dwarfs, neutron stars and core-collapse supernovae. Our
Sun also emits neutrinos, but in this case   the neutrino emission is
a by-product  of nuclear  fusion. For advanced  stages of stellar
evolution, like red giants and  white dwarfs, neutrinos are produced
by thermal effects,  without nuclear reactions being involved\footnote{For
  the Sun, thermal neutrino emission is found to be irrelevant
 \cite{1996slfp.book.....R}.}.  Specifically, neutrino emission in pre-white
dwarf and hot white dwarf stars is the result of  different scattering
processes, with plasmon decay process  $[\gamma \rightarrow
  \overline{\nu} \nu]$ and bremsstrahlung $[e^-(Ze) \rightarrow
  (Ze)e^-\ \overline{\nu} \nu]$ the most relevant  ones.  Plasmon
decay process is an  important neutrino emission mechanism for
a broad range of  temperatures and densities, even though neutrinos do
not couple directly to photons\footnote{Plasmon decay has place owing
  to an \emph{indirect} coupling between neutrinos and photons through
electrons in a plasma \cite{1996slfp.book.....R}.}. It has been
speculated, however, that neutrinos could have non-trivial
electromagnetic properties, in particular a non-zero magnetic dipole
moment ($\mu_\nu$), allowing the plasmon emission process to be much
more efficient. In this case, the non-vanishing magnetic dipole moment
results in a \emph{direct} coupling between neutrinos and the
electromagnetic field (photons). It has been 
shown \cite{1963PhRv..132.1227B} that
it is possible to use the observed properties of stars to constrain
the possible amount of anomalous energy loss and thus the neutrino
electromagnetic properties  \cite{2012arXiv1201.1637R}. 
The neutrino magnetic dipole moment was first calculated in 1980
\cite{1980PhRvL..45..963F}.

For both  pre-white dwarf and hot white dwarf stars, 
neutrino emission is more important than surface photon cooling, and
hence, neutrino losses essentially control the evolutionary  timescale
\cite{1975ApJ...200..306L,1996slfp.book.....R,2004ApJ...602L.109W}. In
particular, plasmon  neutrino emission is the  dominant cooling
mechanism in  hot white dwarfs down to an effective temperature
between $\sim 31\,000$ K and $\sim 23\,000$ K, depending on the value
of the stellar mass.  Since the existence of a neutrino magnetic
dipole moment enhances the plasmon neutrino losses, it is interesting
to investigate the effects that a non-zero
$\mu_\nu$  have on the evolution of white dwarfs. In principle,
this can be done by employing the white dwarf luminosity function
(WDLF). Neutrino cooling causes a depression (the ``neutrino dip'') of
the WDLF at the bright end, which would be enhanced by additional
cooling caused by a magnetic dipole moment, allowing to derive a limit
on $\mu_\nu$.  Employing this approach (see ref. \cite{1976PhRvD..13.2700S}
for the original idea),  it was found  
$\mu_\nu \lesssim 10^{-11} \mu_{\rm B}$ \cite{1994MNRAS.266..289B},  
where $\mu_{\rm B}= e
\hbar/(2 m_e c)$ is the Bohr magneton. The most recent estimate 
of a limit for the neutrino magnetic dipole moment derived from the WDLF is 
$\mu_\nu \lesssim 5 \times 10^{-12} \mu_{\rm B}$ \cite{2014A&A...562A.123M}. 

An alternative way  to constrain the neutrino magnetic dipole moment
from white dwarfs is provided by asteroseismology. Indeed, DB
(pure He atmosphere)  white dwarfs go across a $g$-mode pulsation
instability phase at effective temperatures in the range $29\,000$ K
$\gtrsim T_{\rm eff} \gtrsim 22\,000$ K (the DBV or V777 Her
instability strip \cite{2008ARA&A..46..157W,2008PASP..120.1043F,review}),
Their evolutionary timescale can be
inferred, in principle, by measuring the temporal rates of period changes
($\dot{\Pi}\equiv d\Pi/dt$,  $\Pi$ being the pulsation
period). Interestingly enough,  the range of effective temperatures in
which the evolution of DB white dwarfs  is dominated by neutrino
emission partly overlaps with the DBV instability strip,  and thus,
the magnitude of the rates of period change of DBV stars 
is set by plasmon neutrino emission.
Therefore, if a non-zero neutrino magnetic dipole moment exists, the
value of $\dot{\Pi}$ should be enhanced in comparison with the case in
which $\mu_\nu= 0$, and this  provides a bound to
$\mu_\nu$. A similar approach has been used in the context of
pulsating DA (H-rich atmospheres)  white dwarfs to derive 
bounds on the axion mass 
\cite{Isern92,2001NewA....6..197C,2008ApJ...675.1512B,2012MNRAS.424.2792C,2012JCAP...12..010C}, and also to place
constraints on  the secular rate of variation of the gravitational
constant $G$  \cite{2013JCAP...06..032C}. For DBV stars, no accurate
measurement of the rate of period change has been yet derived for any
pulsating star. However, it has been possible to determine a preliminary
value of $\dot{\Pi}= (2.0 \pm 0.9) \times 10^{-13}$ s/s 
for the largest amplitude mode with period at $\sim 489$ s 
of the pulsating star  PG 1351+489, the first such estimate for 
a DBV \cite{2011MNRAS.415.1220R}.
According to its spectroscopic effective temperature ($22\,000$ K $- 26\,000$
K),  this star could  be too cool  to allow a measurement of  the
\emph{normal} plasmon emission rate.  However, it can be used for 
detecting \emph{anomalous}  neutrino loss.

In this paper, we employ the above mentioned estimate of the rate of 
period change for PG 1351+489 to derive an
upper limit to the neutrino magnetic dipole moment.  The paper is
organized  as follows. In section \ref{models} we describe the input
physics employed  in the computation of our evolutionary DB white
dwarf models,  and the effects that a non-zero $\mu_\nu$ have on the DB
white dwarf evolution.  Section \ref{effects} is devoted to describe
the impact of $\mu_\nu$  on the pulsation properties of DB white
dwarfs.  In section \ref{constraints} we derive constraints on the
neutrino magnetic dipole moment by considering the rate of period
change estimate for the $\sim 489$ s period of the star PG 1351+489 by
relying on their spectroscopically determined effective temperature
alone. In section \ref{astero} we obtain an upper limit on $\mu_\nu$ by
employing  an asteroseismological model for PG 1351+489. Finally, in
section \ref{conclusions}, we compare our results with current
astrophysical bounds on $\mu_\nu$, and present our concluding remarks.

\section{DB white dwarf models}
\label{models}

\begin{figure}
\centering
\includegraphics[clip, width=0.7\textwidth, angle=0]{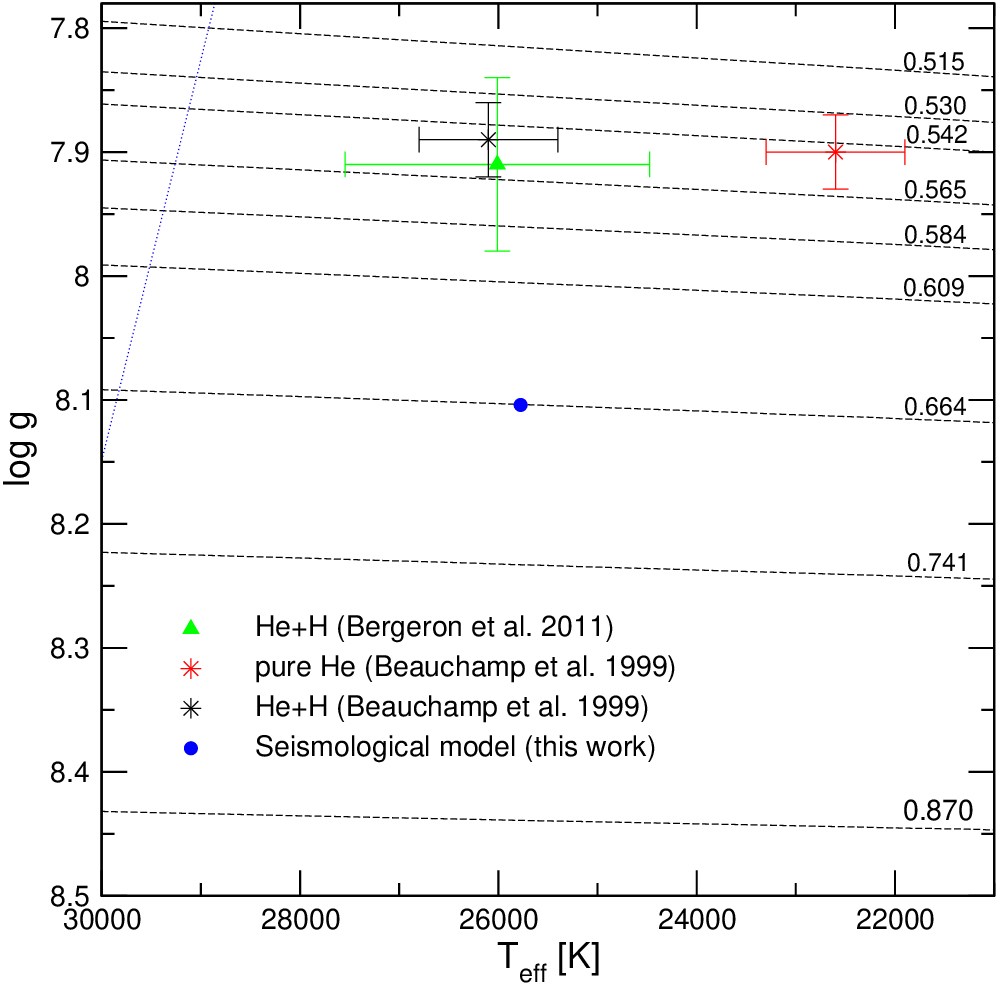}
\caption{Our set of DB white dwarf evolutionary tracks on the $T_{\rm eff} - 
\log g$ plane, labeled with the values of the stellar mass. 
The location of PG 1351+489 \cite{1999ApJ...516..887B} 
is emphasized with a red symbol (a black  symbol)
assuming a pure He atmosphere  
(considering a He rich atmosphere with impurities of H). 
The location of the star as given by Ref. \cite{2011ApJ...737...28B}
(atmosphere rich in He plus traces of H) is shown with a green triangle. 
Finally, the location of the star as predicted by 
our asteroseismological model (section \ref{astero}) is 
marked with a blue circle. The theoretical blue
edge of the DBV instability strip  is depicted with a blue 
dashed line \cite{2009JPhCS.172a2075C}. See the o-line edition 
of the journal for a colo version of this figure.}
\label{hr}
\end{figure}

The  evolutionary DB  white dwarf  models employed  in this  work were
computed using the {\tt  LPCODE} stellar
evolutionary code \cite{2005A&A...435..631A,2009ApJ...704.1605A}.
All  our white dwarf
sequences were computed in a  consistent way with the evolution of the
chemical abundance distribution due to element diffusion along the
entire  cooling  phase.   Neutrino emission rates for
pair, photo,  and bremsstrahlung  processes are taken 
into account \cite{1996ApJS..102..411I}.  
For plasma  processes, we used the treatment presented in Ref. 
\cite{1994ApJ...425..222H}.  Of particular interest in this paper is the 
anomalous energy loss due to the existence
of a magnetic dipole moment, $\epsilon_\nu^{\rm mdm}$. It was computed 
from the plasmon neutrino emission, $\epsilon_\nu^{\rm p}$, using 
the scaling relation \cite{1994ApJ...425..222H}:

\begin{equation}
 \epsilon_\nu^{\rm mdm}= 0.318\ \mu_{\rm 12}^2 
\left(\frac{10\; {\rm keV}}{\hbar \omega_{\rm p}}\right)^2 
\frac{Q_2}{Q_3}\ \epsilon_\nu^{p}
\end{equation}

\noindent Here, $\mu_{12}= \mu_\nu 
/\left(10^{12} \mu_{\rm B} \right)$, $Q_2/Q_3 \sim 1$ 
\cite{1994ApJ...425..222H},
and $\omega_{\rm p}$, the plasma frequency, was computed as 
in \cite{1996slfp.book.....R} (see Ref. \cite{2014A&A...562A.123M} 
for details).

The  initial models  for our  DB white  dwarf sequences  correspond to
realistic  H-defficient PG 1159  stars (the precursors of H-defficient 
white dwarfs) derived  from  the  full
evolutionary calculations of  their progenitor stars 
\cite{2005A&A...435..631A,2006A&A...454..845M}. All the  sequences  were 
computed  from the Zero Age Main Sequence through the  
thermally-pulsing and mass-loss  phases on the  
Asymptotic Giant Branch and finally to  the born-again  stage where the  
remaining H  is violently burned.   After  the born  again  episode,  
the H-deficient,  quiescent He-burning  remnants evolve at  
constant luminosity  to the  domain of PG 1159 stars with  a surface 
chemical composition rich in  He, C and O \cite{2006A&A...454..845M}
and eventually to the DB white dwarf stage. This set  of  PG 1159 
evolutionary  models has  succeeded  in 
explaining  both  the spread  in surface chemical  composition 
observed in  most PG 1159 stars  and the
location of  the GW Vir instability  strip in the 
$\log  T_{\rm eff}- \log g$  plane \cite{2006A&A...458..259C}.   
Also, these PG  1159 models
have  been employed  in  detailed asteroseismological  studies of  seven
pulsating  PG 1159  stars  \cite{2007A&A...461.1095C,2007A&A...475..619C,
2008A&A...478..869C,2009A&A...499..257C,2014arXiv1405.5075K}. 
It is worth remarking that realistic initial white dwarf stuctures
are needed in order to accurately predict the evolutionary 
properties of DBVs. In fact,  it has been 
shown \cite{2013A&A...555A..96S} that differences larger than $10\%$
in the cooling times may arise from assuming different thermal structures
of the first white dwarf converged models at the beginning of the 
cooling sequences. 

\begin{figure}
\centering
\includegraphics[clip, width=0.7\textwidth, angle=0]{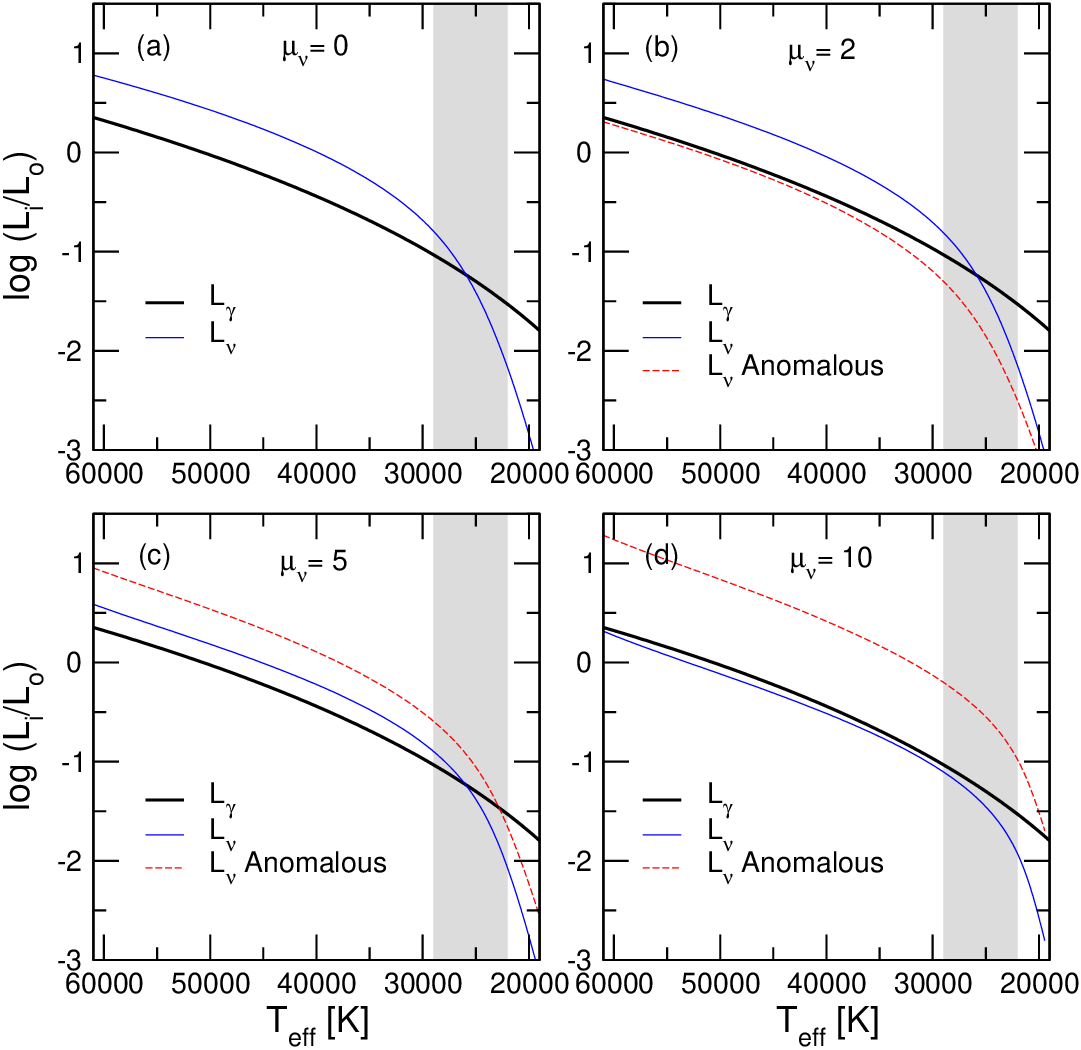}
\caption{Luminosity contributions 
for a $0.664 M_{\odot}$ DB white-dwarf star in terms of the 
effective temperature. Panel (a): photon
luminosity ($L_{\gamma}$), and neutrino losses without a 
magnetic dipole moment ($L_{\nu}$). Panels (b), (c), and (d): 
neutrino emission due to standard model processes ($L_{\nu}$) but 
assuming the existence of a neutrino magnetic 
dipole moment $\mu_\nu$ (in units of $10^{-12} \mu_{\rm B}$), and 
anomalous neutrino emission ($L_{\nu}$ Anomalous). 
The gray area indicates the domain of DBV pulsating stars.}
\label{lumis}
\end{figure}

Specifically, we considered nine DB white dwarf sequences with stellar
masses:  $0.515$, $0.530, 0.542, 0.565, 0.584,  0.609, 0.664, 0.741$,
and $0.870  M_{\odot}$.  These DB  sequences are characterized  by the
maximum He-rich  envelope that  can be left  by the previous evolution  
if it is assumed that they are the result  of a born-again episode. 
The value of the
envelope  mass ranges  from  $M_{\rm He}/M_*  \sim  2 \times  10^{-2}$
($M_*=  0.515 M_{\odot}$) to  $M_{\rm He}/M_*  \sim 1  \times 10^{-3}$
($M_*=  0.870  M_{\odot}$).   The  complete  set  of  DB  white-dwarf
evolutionary sequences (computed neglecting a neutrino magnetic 
dipole moment) is displayed in the $T_{\rm eff} - 
\log g$ diagram of Fig.   \ref{hr}, where $g$ is the 
surface gravity defined
as $g= G M_* / R_*^2$, $G$ being the gravitational constant 
and $R_*$ the stellar radius. Also shown  
is the spectroscopic location of PG 1351+489
and the location of our asteroseismological model for this star 
(section \ref{astero}).

The impact of a non-zero neutrino magnetic dipole moment on the 
evolution of white dwarfs has been explored in detail 
\cite{2014A&A...562A.123M}. Here we briefly show how our DB white dwarf
evolutionary sequences are affected when a 
non-vanishing neutrino magnetic dipole moment 
is taken into account. The initial 
DB models of our sequences were computed under the standard assumptions, 
that is, neglecting anomalous neutrino losses. We swiched on 
the anomalous neutrino losses at very high effective
temperatures ($T_{\rm eff} \sim 70\,000$ K) in such a way that the unphysical
transitory associated with this artificial procedure finishes long before the
models reach the instability strip of DBV stars ($29\,000$ K $\gtrsim 
T_{\rm eff} \gtrsim 22\,000$ K). Apart from the standard sequences
(that assume $\mu_\nu= 0$), we computed 
additional white dwarf sequences considering a non-zero magnetic 
dipole moment. The effect of adopting $\mu_\nu \neq 0$ is depicted in 
Fig. \ref{lumis}, in which we 
show the luminosity contributions for the case of 
a DB white-dwarf star with $M_*= 0.664 M_{\odot}$ in terms of the 
effective temperature. In panel (a) we show the standard case in which
anomalous neutrino emission is neglected. Photon luminosity ($L_{\gamma}$)
and neutrino losses ($L_{\nu}$) without a magnetic dipole moment 
($\mu_\nu= 0$) are displayed with a thick black curve and a thin blue curve, 
respectively. In panels (b), (c), and (d), we show the cases in which
we allow neutrino emission due to standard model processes 
(again, $L_{\nu}$) but considering the existence of a neutrino magnetic 
dipole moment $\mu_\nu$. In these panels, 
we include also anomalous neutrino emission ($L_{\nu}$ Anomalous),
depicted with thin dashed red curves.
The assumed values of $\mu_\nu/\left(10^{12} \mu_{\rm B} \right)$
are 2, 5, and 10. The gray area indicates the domain of DBV pulsating stars.
Fig. \ref{lumis} shows that as the anomalous neutrino
emission  is increased by increasing
$\mu_\nu$, the feedback on the thermal structure of the white dwarf
forces neutrino emission through the standard channels to be lower.
In particular, for $\mu_\nu/\left(10^{12} \mu_{\rm B} \right) \gtrsim 3$
(not shown), the anomalous neutrino losses overcome neutrino emission due to 
standard model processes. The net effect on the evolution is that 
the higher the values of $\mu_\nu$, the faster the cooling of the white 
dwarf \cite{2014A&A...562A.123M}. 
In summary, the inclusion of a neutrino magnetic dipole moment 
in the evolution
of white dwarfs gives rise to an additional cooling mechanism. This 
effect is identical to that found when axion emission is considered
in the evolution of cooler white dwarfs \cite{Isern92}.

\section{The effects of $\mu_\nu$ on the pulsations of white dwarfs}
\label{effects}

Here we explore the effects that a neutrino magnetic dipole 
moment $\mu_\nu$ has on the pulsation periods and rates of period change 
of white dwarf stars. The  pulsation periods  employed in  the present 
analysis were
computed with the pulsational code {\tt LP-PUL} 
\cite{2006A&A...454..863C}. This code has been employed in 
numerous pulsational studies of white dwarfs 
(see Ref. \cite{review} and references therein). Pulsations in 
white dwarfs are associated to non-radial $g$(gravity)-modes 
which are a sub-class of spheroidal modes\footnote{Spheroidal modes 
are characterized by $(\vec{\nabla} \times \vec{\xi})_r= 0$
and $\sigma \neq 0$, where $\vec{\xi}$ is the 
vector Lagrangian displacement and $\sigma$ the pulsation frequency 
\cite{1989nos..book.....U}.} whose main restoring force 
is gravity. These modes are characterized by low oscillation frequencies
(long periods) and by a displacement of the stellar fluid essentially 
in the horizontal direction. $g$-modes are labelled 
with the harmonic degree $\ell = 0, 1, 2, \cdots, \infty$ (the number of nodal 
lines in the stellar surface), the azimuthal order 
$m= 0, \pm 1, \cdots, \pm \ell$ (the number of such nodal lines 
in longitude), and the radial order $k= 0, 1, 2,\cdots, \infty$ 
(the number of nodes in the radial component of the eigenfunction) 
\cite{1989nos..book.....U,2008ARA&A..46..157W,review}.

\begin{figure}
\centering
\includegraphics[clip, width=1.0\textwidth, angle=0]{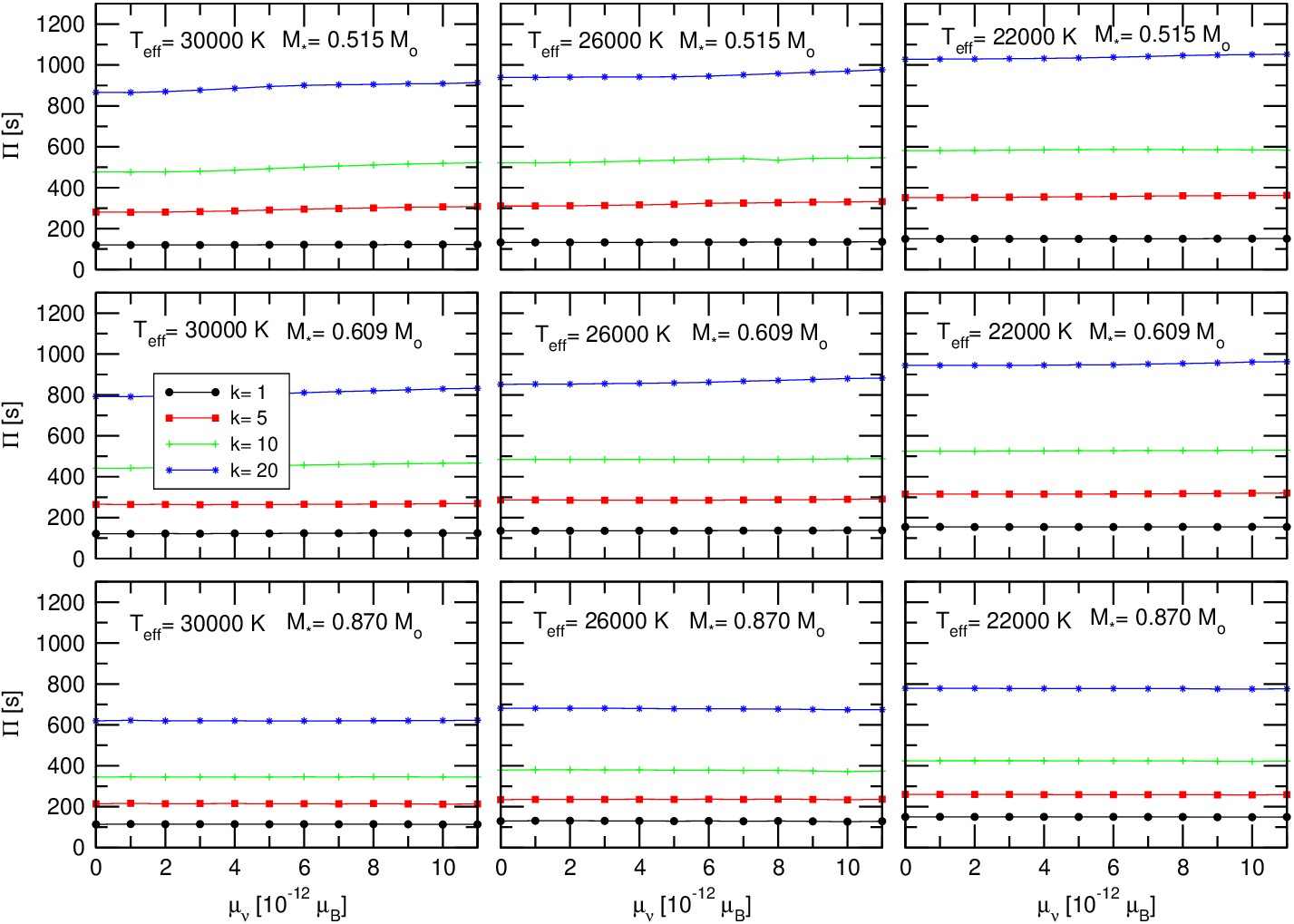}
\caption{The pulsation periods of $\ell= 1$ $g$-modes with 
radial order $k= 1, 5, 10$ and 20 for DB white dwarf models
with masses $M_*/M_{\odot}= 0.515, 0.609$ and 0.870, and
effective temperatures $T_{\rm eff} \sim 30\,000, 26\,000$ and $22\,000$ K,
in terms of the neutrino magnetic dipole 
moment.}
\label{p-mu}
\end{figure}

\begin{figure}
\centering
\includegraphics[clip, width=1.0\textwidth, angle=0]{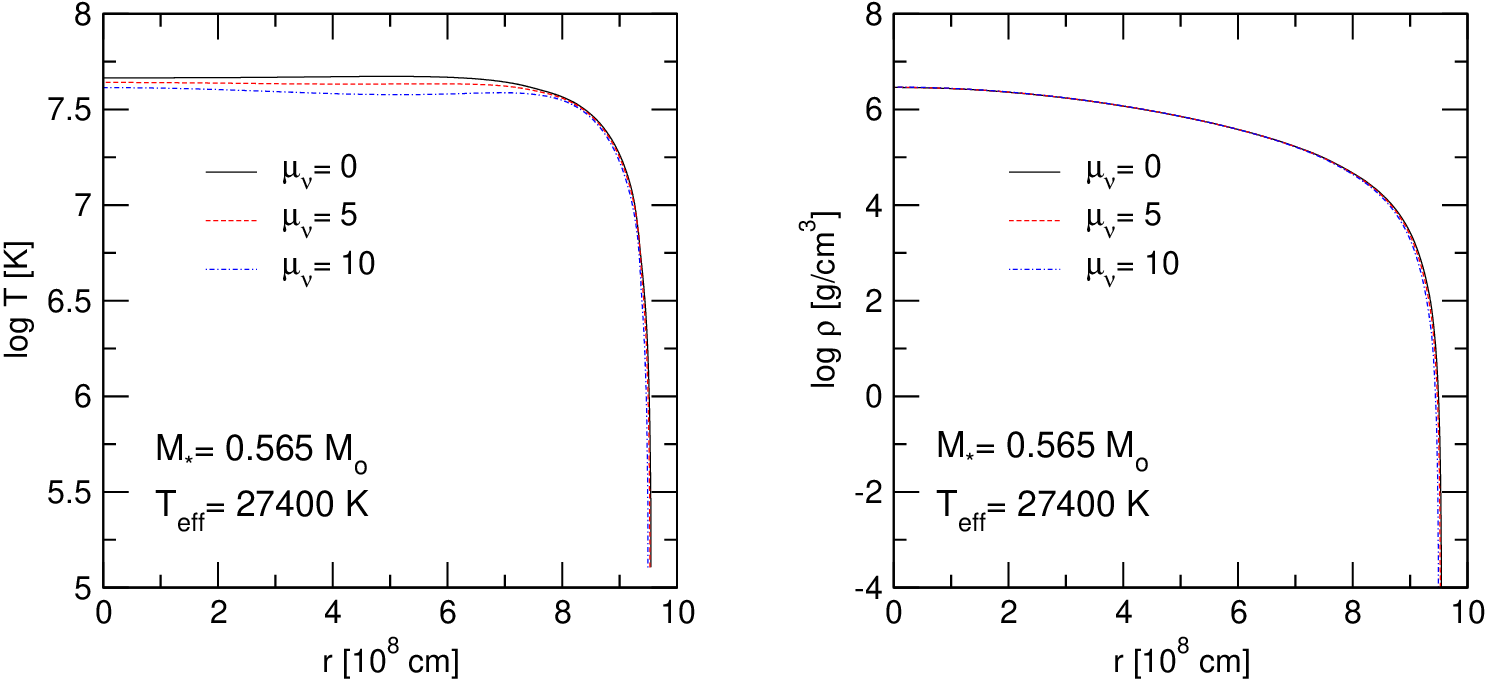}
\caption{ The logarithm of the temperature (left panel) and density 
(right panel) profiles  
corresponding to a DB white dwarf model with $M_*= 0.565 M_{\odot}$ 
and $T_{\rm eff}= 27\,400$ K for the cases of $\mu_{\nu}= 0, 5$ and $10$ 
(in units of $10^{-12} \mu_{\rm B}$).}
\label{r-t-rho}
\end{figure}

We have computed  dipole and quadrupole 
($\ell= 1$ and  $\ell= 2$) $g$-modes with pulsation periods in the range 
$100$ s $\lesssim \Pi \lesssim 1200$ s, thus covering the 
range of observed periods in DBV white dwarfs. We 
considered DB white dwarf models with masses 
in the range $0.515 \lesssim M_*/M_{\odot} \lesssim 0.870$
and effective temperatures in the interval $30\,000$ K $\gtrsim
T_{\rm eff} \gtrsim 20\,000$ K, thus embracing the mass and effective 
temperature ranges in which DBV white dwarfs are observed. 
In order to assess the impact of a non-zero neutrino magnetic dipole moment 
on the periods and temporal rates of period 
changes, we carried out these 
computations for model sequences in which we vary $\mu_\nu$
in the range $0 \leq \mu_\nu/\left(10^{12} \mu_{\rm B} \right) \leq 11$.

\begin{figure}
\centering
\includegraphics[clip, width=1.0\textwidth, angle=0]{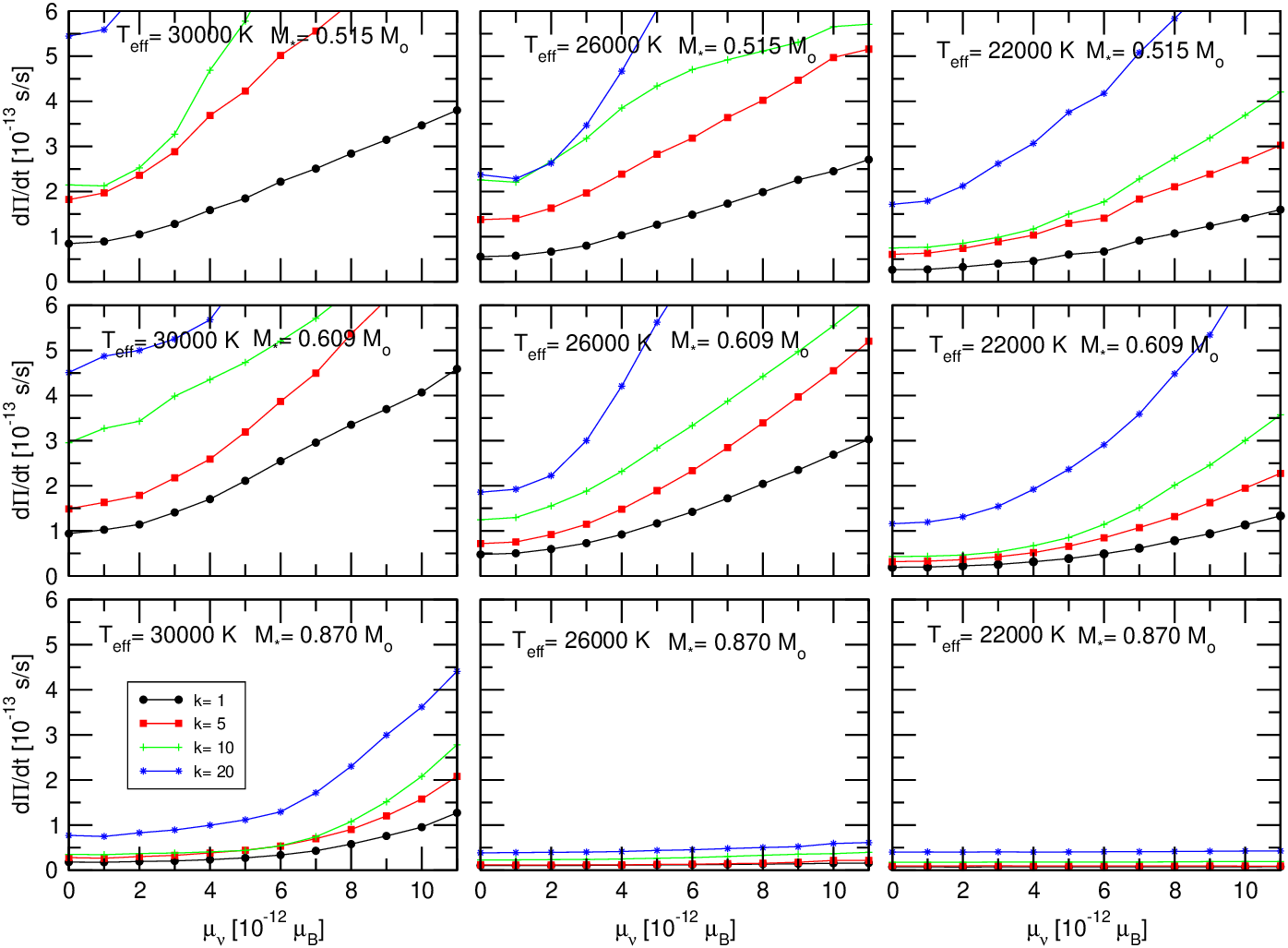}
\caption{Same as in Fig. \ref{p-mu}, but for the rates of period 
change.}
\label{dpdt-mu}
\end{figure}

In Fig. \ref{p-mu} we show the dipole pulsation periods 
with $k= 1, 5, 10$ and 20 ($100$ s $\lesssim \Pi \lesssim 1000$ s) 
for DB white dwarf models with masses 
$M_*/M_{\odot}= 0.515, 0.609$ and 0.870, and effective temperatures 
$T_{\rm eff} \sim 30\,000, 26\,000$ and $22\,000$ K,
in terms of the neutrino magnetic dipole moment ($\mu_\nu$).
The variation in the periods is very small, in spite of the rather 
wide range of $\mu_\nu$ considered. This result, which was 
first found in the case of axion emission \cite{2001NewA....6..197C,
2012MNRAS.424.2792C,2012JCAP...12..010C},
implies that due to the additional cooling mechanism induced 
by the existence of a neutrino magnetic dipole moment,
the structure of the white dwarf models themselves are slightly 
affected. Indeed, in Fig. \ref{r-t-rho} we show the logarithm 
of the temperature and density profiles of a DB white dwarf model with
$M_*= 0.565 M_{\odot}$ and $T_{\rm eff}= 27\,400$ K for  
different values of $\mu_{\nu}$, including the case $\mu_{\nu}= 0$.
The effects of a neutrino magnetic dipole moment on these structural 
quantities are barely seen in the plots. As a result of this,  
for a fixed value of the effective
temperature, the pulsation periods are largely independent of the
adopted value for $\mu_\nu$. The impact of a non-zero neutrino magnetic 
dipole moment on the rates of period changes is depicted in 
Fig. \ref{dpdt-mu}, where we show $\dot{\Pi}$ for the same pulsation 
modes and the same white dwarf models shown in Fig. \ref{p-mu} in terms
of $\mu_\nu$. At variance with what happens with the pulsation
periods in most cases, the values of $\dot{\Pi}$ are strongly affected by the 
additional cooling source, substantially increasing for increasing 
values of $\mu_\nu$. The effect is stronger for high effective 
temperatures and low stellar masses ($M_*= 0.515 M_{\odot}$, 
$T_{\rm eff}= 30\,000$ K), and is barely noticeable for
low effective temperatures and high masses ($M_*= 0.870 M_{\odot}$, 
$T_{\rm eff}= 22\,000$ K). 

\section{Constraints on $\mu_\nu$ from the DBV star PG 1351+489}
\label{constraints}

The star PG 1351+489 is one of the 21 pulsating DB (He-rich atmosphere)
white dwarfs known up to date \cite{2011ApJ...736L..39O}. 
Since the discovery of pulsations in this star \cite{1987ApJ...316..305W},
it was realized that it could be a candidate for the 
first measurement of a rate of period change in a DBV star, 
because its power spectrum is dominated by a single high-amplitude
pulsation mode with a period at $\sim 489$ s. In addition,  two smaller 
amplitude modes and several linear combination frequencies 
were detected \cite{2003BaltA..12...33A}. 
The pulsations in this star were re-analyzed, and the previous results
were confirmed, providing more precise periods and also an additional 
low-amplitude mode \cite{2011MNRAS.415.1220R}. More importantly, 
it was possible to obtain, for the first time for a DBV star, an estimate of 
the rate of period change for the period at $\sim 489$ s of 
$\dot{\Pi}= (2.0\pm 0.9) \times 10^{-13}$ s/s \cite{2011MNRAS.415.1220R}. 
Although not definitive, this estimate constitutes a
very valuable opportunity to place constraints on 
$\mu_\nu$ \emph{independently} from other techniques.
In fact, we can compare the estimated value of the rate of period change
for the mode with period $\sim 489$ s of PG 1351+489 with theoretical 
values of the rate of period changes computed under the 
assumption of a non-zero $\mu_\nu$. The value of $\mu_\nu$
for which the theoretical $\dot{\Pi}$ matches the observed
rate of period change, may be regarded as an estimation of the 
magnitude of $\mu_\nu$. In order to apply such a procedure, we have to choose
a DB white dwarf model with an effective temperature
close to the observed one. Also, we have to consider
theoretical values of $\dot{\Pi}$ corresponding to 
modes with periods 
close to the observed period ($\sim 489$ s). Regarding the effective 
temperature and gravity of PG 1351+489, there are several spectroscopic 
determinations  for this star. The first detailed analysis
\cite{1999ApJ...516..887B} provided $T_{\rm eff}= 22\,600 \pm 700$ K, 
$\log g= 7.90 \pm 0.03$ using pure He atmospheres, and 
$T_{\rm eff}= 26\,100 \pm 700$ K, $\log g= 7.89 \pm 0.03$
employing atmospheres with impurities of H. 
The most recent analysis \cite{2011ApJ...737...28B}
indicates that $T_{\rm eff}= 26\,010 \pm 1536$ K, $\log g= 7.91 \pm 0.07$, 
obtained by assuming H contamination 
($\log {\rm H}/{\rm He}= −4.37 \pm 0.82$). 

\begin{figure}
\centering
\includegraphics[clip, width=0.7\textwidth, angle=0]{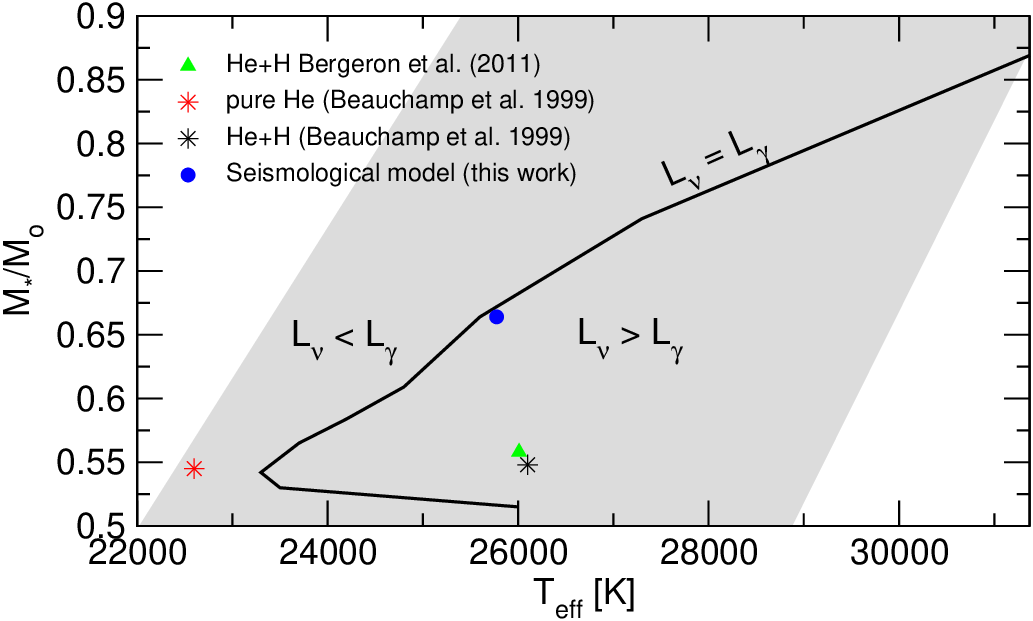}
\caption{A $M_*-T_{\rm eff}$ diagram showing the regions 
in which neutrino luminosity dominates over photon luminosity 
($L_\nu > L_\gamma$), and the regions in which the opposite holds 
($L_\nu < L_\gamma$). The black curve corresponds to the case
in which $L_\nu= L_\gamma$. The location of the star PG 1351+489 is 
displayed according to different spectroscopic determinations 
and also as predicted by an asteroseismological model 
(section \ref{astero}). The gray band indicates the DBV instability strip.}
\label{neutri-domain}
\end{figure}

According to the spectroscopic effective temperature of PG 1351+489
($22\,000$ K $- 26\,000$ K),  it could be argued that this star
is a bit cool for studying plasmon emission processes 
\cite{2011MNRAS.415.1220R}.
In Fig. \ref{neutri-domain} we show the regions in which
neutrino luminosity dominates over photon luminosity (to the right of
the black curve),  and the regions in which photon luminosity
overwhelms neutrino luminosity (to the left of the black curve). We
have included the location of PG 1351+489 on this  plane according to
different estimations of the $T_{\rm eff}$ and $M_*$. As can be seen,
in all of the cases the star is located in
the region in which neutrino emission  dominates. From this plot, we
conclude that the star is actually an appropriate  target to study
neutrino emission, and in particular, a hypotetical anomalous 
neutrino emission. 

\begin{figure}
\centering
\includegraphics[clip, width=0.7\textwidth, angle=0]{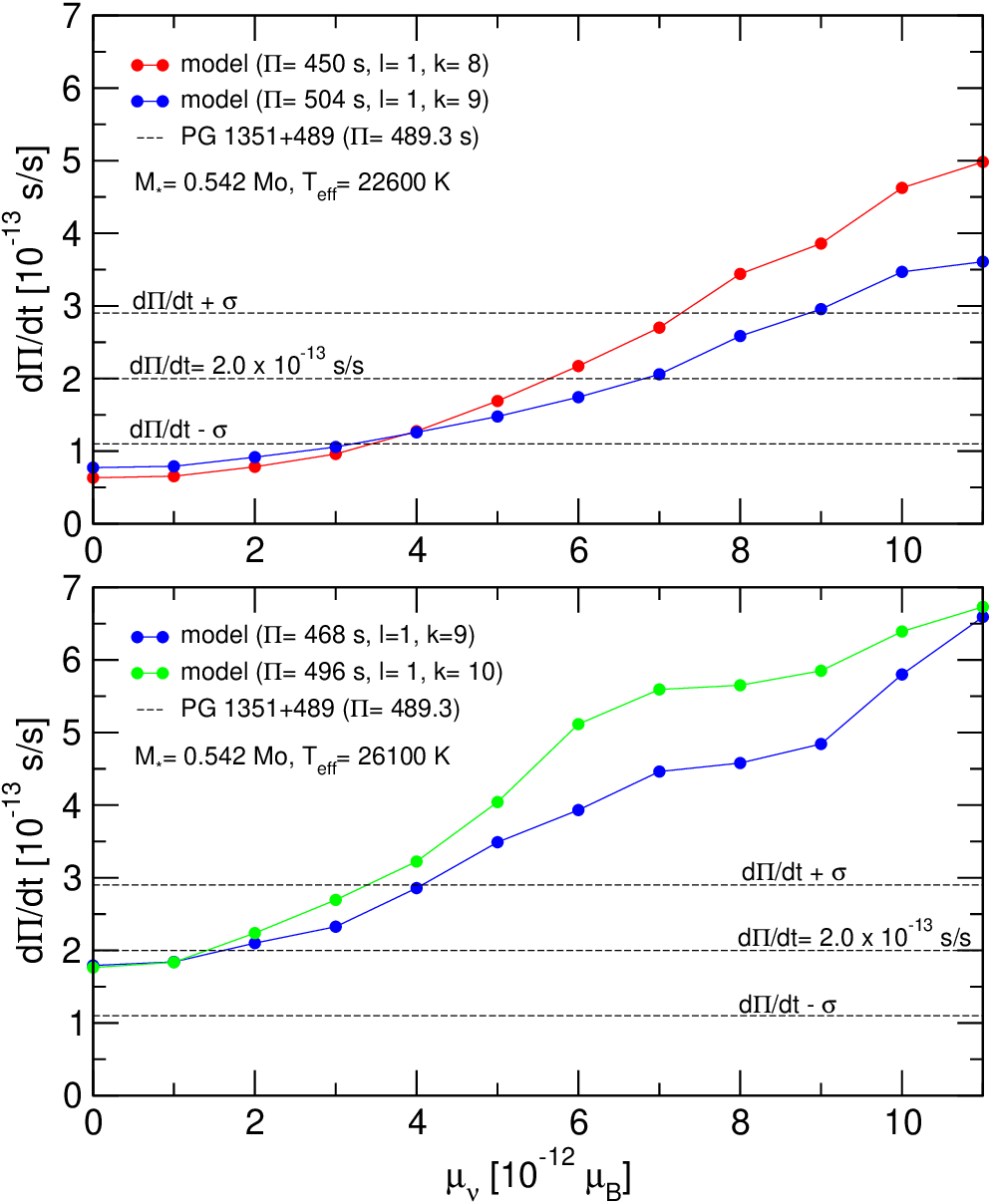}
\caption{The rates of period change of selected modes in terms
of the neutrino magnetic dipole moment, corresponding to 
models with $M_*= 0.542 M_{\odot}$ and $T_{\rm eff}= 22\,600$ K
(upper panel), and  $T_{\rm eff}= 22\,600$ K (lower panel).
The estimate of  the rate of period 
change of the 489 s period \cite{2011MNRAS.415.1220R} of 
PG 1351+489 and its uncertainties are shown using thin dashed 
horizontal lines.}
\label{dpdt-mu-0542}
\end{figure}

\begin{figure}
\centering
\includegraphics[clip, width=0.7\textwidth, angle=0]{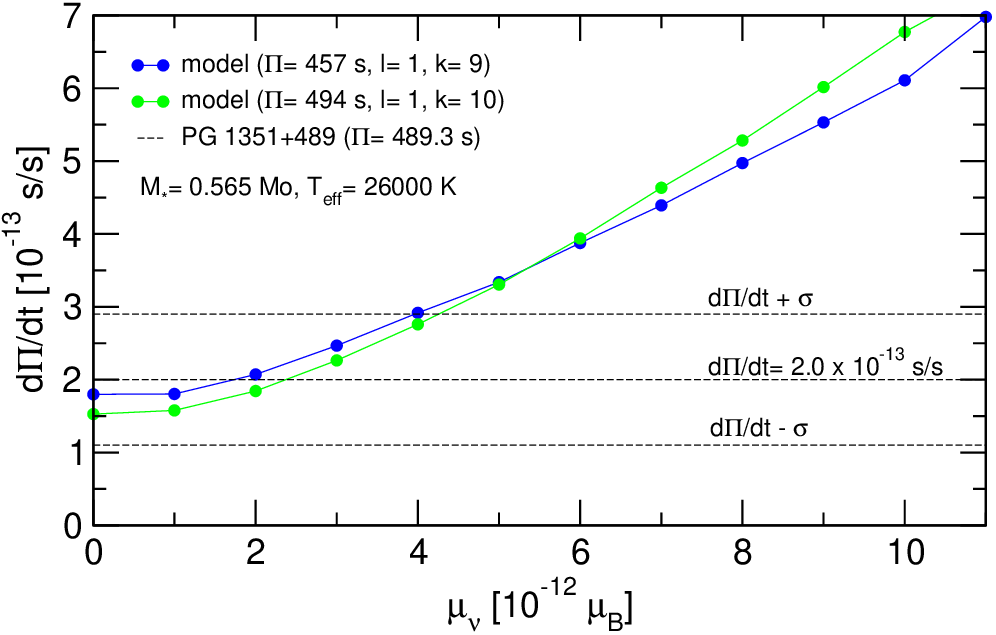}
\caption{Same as Fig. \ref{dpdt-mu-0542}, but for the case
of a model with a stellar mass of $M_*= 0.565 M_{\odot}$ and 
$T_{\rm eff}= 26\,000$ K. }
\label{dpdt-mu-0565}
\end{figure}

Following  the spectroscopic analysis of Ref. \cite{1999ApJ...516..887B}, 
the stellar mass of PG 1351+489
must be $M_* \sim 0.542 M_{\odot}$ according to our evolutionary 
tracks (see Fig. \ref{hr}). Accordingly, we have considered two DB white dwarf
models belonging to this sequence, one of them with 
$T_{\rm eff} \sim  22\,600$ K and the other 
one  with $T_{\rm eff} \sim  26\,100$ K, in order to include
the case in which the atmosphere is made of pure He, and also
the case in which there are  small abundances of H. 
In Fig. \ref{dpdt-mu-0542}
we show the theoretical rates of period changes 
in terms of $\mu_\nu$ for dipole modes with periods of $\sim 450$ s ($k= 8$) and
$\sim 504$ s ($k= 9$) corresponding 
to the model with $T_{\rm eff}= 22\,600$ K (upper panel) and modes with periods 
of $\sim 468$ s ($k= 9$) and $496$ s ($k= 10$) 
corresponding to the  model with $T_{\rm eff}= 26\,100$ K (lower panel).
Note that in both cases the periods adopted embrace the observed
period ($\sim 489$ s). By comparing the theoretical values of $\dot{\Pi}$ 
with the estimate of $\dot{\Pi}$ for PG 1351+489 
($\dot{\Pi}= (2.0\pm 0.9) \times 10^{-13}$ s/s), we infer 
$3 \lesssim \mu_\nu /\left(10^{12} \mu_{\rm B}\right)  
\lesssim 9$ in the first case, and $\mu_\nu /\left(10^{12} \mu_{\rm B}\right)  
\lesssim 4$ in the second case. We repeated this analysis
by considering the modern estimate for $T_{\rm eff}$
and $\log g$ \cite{2011ApJ...737...28B}. In this case, according 
to our evolutionary tracks, the stellar mass of PG 1351+489 
is $M_* \sim 0.565 M_{\odot}$ (see Fig. \ref{hr}).
The results are shown in Fig. \ref{dpdt-mu-0565}. The upper bound 
in this case is $\mu_\nu /\left(10^{12} \mu_{\rm B}\right)  
\lesssim 4$, in perfect agreement with the upper limit derived 
by adopting the high effective temperature determination 
of \cite{1999ApJ...516..887B}. 

\section{Constraints on $\mu_\nu$ from an asteroseismological 
model for PG 1351+489}
\label{astero}

Another way in which we can derive a constraint on the neutrino 
magnetic dipole moment is by employing a seismological model for 
PG 1351+489, that is, the DB white dwarf model that best reproduces 
the observed periods exhibited by the star. In obtaining the 
asteroseismological model, we neglect the existence of a 
neutrino magnetic dipole moment, that is, we assume $\mu_\nu= 0$,
since we have shown that the period pattern is not affected by the 
inclusion of $\mu_\nu$. In order to search for the asteroseismological 
model, we employ a quality function:

\begin{equation}
\chi^2(M_*, T_{\rm eff})= \frac{1}{N} \sum_{i=1}^{N} 
\min[(\Pi_{{\rm obs},i}- \Pi_k)^2], 
\end{equation}

\noindent that measures the  goodness of  the  match between  the
theoretical  pulsation periods ($\Pi_k$)  and the  observed individual
periods  ($\Pi_{{\rm obs},  i}$). Here, $N$  ($= 4$) is the number of  
observed periods, which are: $\Pi_{{\rm obs},i}= 335.26, 489.33, 584.68$
and $639.63$ s \cite{2011MNRAS.415.1220R}. The DB
white dwarf model  that shows the lowest value  of $\chi^2$ is adopted
as the ``best-fit model''.  We have considered $\ell= 1$ and 
$\ell= 2$ modes. The quality of our period fits 
is assessed by means of the  average of the absolute  period 
differences, $\overline{\delta}=  (\sum_{i=1}^N  |\delta_i|)/N$,  
where  $\delta_i=
\Pi_{{\rm  obs}, i}  -\Pi_k$,  by  the root-mean-square  residual,
$\sigma= \sqrt{(\sum |\delta_i|^2)/N}= \sqrt{\chi^2}$, and by 
the  Bayes Information  Criterion 
${\rm BIC}= N_{\rm p} \left(\log N/N \right) + \log \sigma^2$,
where $N_{\rm p}$ is  the number of free parameters, and $N$
the number of  observed periods \cite{2000MNRAS.311..636K}. 
In our case,  $N_{\rm p}= 2$ (stellar
mass and  effective temperature).  The  smaller the value of  BIC, the
better the quality  of the fit. 

\begin{figure}
\centering
\includegraphics[clip, width=0.7\textwidth, angle=0]{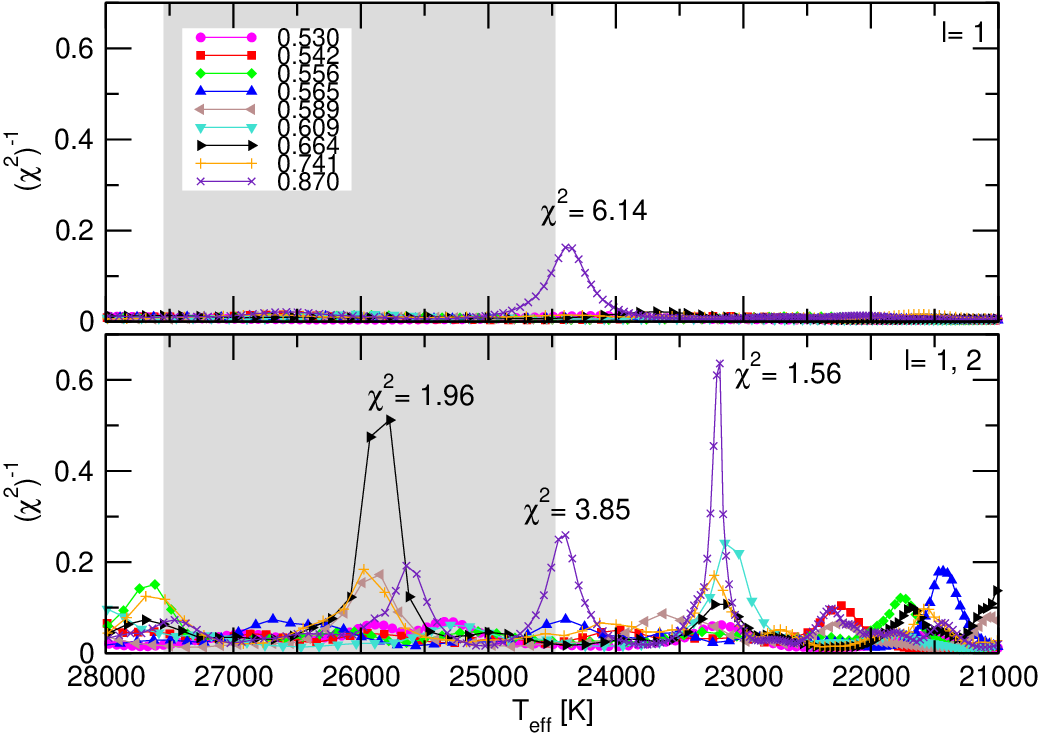}
\caption{The quantity $(\chi^2)^{-1}$ in terms of the effective temperature for
different stellar masses, along with the spectroscopic effective temperature 
of PG 1351+489 and its uncertainties (gray area) according to 
Ref. \cite{2011ApJ...737...28B}.  
The upper panel corresponds to the case in which all the 
periods of PG 1351+489  are assumed  to be $\ell= 1$, and the lower panel 
shows the results for the case in which we allow the observed periods
to be a combination of $\ell= 1$ and $\ell= 2$.}
\label{period-fits}
\end{figure}

\begin{table}
\centering
\caption{The asteroseismological solutions for PG 1351+489 when 
we assume that all
the periods are associated to $\ell= 1$ modes (first row), and in the 
case in which we 
suppose that they are a combination of $\ell= 1$ and $\ell= 2$ modes 
(second to fourth row).}
\begin{tabular}{ccccccc}
\hline
\hline
\noalign{\smallskip}
Solution & $M_*$             & $T_{\rm eff}$ & $\ell$ &$\overline{\delta}$ & $\sigma$ & BIC\\
\noalign{\smallskip}
        &    $[M_{\odot}]$    &   $[$K$]$    &        &  $[$s$]$         &  $[$s$]$ &    \\      
\noalign{\smallskip}
\hline
\noalign{\smallskip}
1       &     0.870          &   $24\,395$  & 1    & 2.167             &   2.478  &  1.089 \\ 
2       &     0.870          &   $23\,184$  & 1,2  & 1.210             &   1.253  &  0.496 \\   
3       &     0.870          &   $24\,395$  & 1,2  & 1.575             &   1.964  &  0.887 \\  
4       &     0.664          &   $25\,775$  & 1,2  & 1.065             &   1.397  &  0.591 \\  
\noalign{\smallskip}
\hline
\hline
\end{tabular}
\label{table1}
\end{table}

The quantity $(\chi^2)^{-1}$ in  terms of the effective temperature
for different stellar masses is shown in Fig. \ref{period-fits}
together with the most recent spectroscopic determination of the
effective temperature of PG 1351+489  and its uncertainties (gray
strip) \cite{2011ApJ...737...28B}.   In the upper panel we depict the
results of the case in which all the periods of PG 1351+489  are
supposed  to be $\ell= 1$. In this case, we found a single solution
corresponding to a best-fit model with $M_*= 0.870 \pm 0.015 M_{\odot}$ 
and $T_{\rm eff}= 24\,395 \pm 180$ K.  For  this  model,  we  obtain
$\overline{\delta}= 2.17$ s, $\sigma=  2.48$ s and   BIC= 1.089. The
effective temperature  of this model  is at the cool boundary of the
interval in $T_{\rm eff}$  allowed by spectroscopy for this star. The
case in which we  allow the modes to be $\ell= 1$ or $\ell= 2$ is
shown in the lower panel of Fig. \ref{period-fits}.  We found three
possible seismological solutions. Two of them  correspond to models
with $M_*= 0.870 M_{\odot}$ and  $T_{\rm eff}= 24\,395$ K and $T_{\rm
  eff}= 23\,184$ K. The third  solution corresponds to a model with
$M_*= 0.664 M_{\odot}$ and  $T_{\rm eff}= 25\,775$ K. In Table
\ref{table1} we summarize the main characteristics of all the
solutions  with the corresponding $\overline{\delta}$, $\sigma$ and
BIC values. Clearly,  when we allow for the observed periods to be a
combination of  $\ell= 1$ and $\ell= 2$ modes, the seismological
solutions are substantially better (solutions 2 to 4) than assuming
that all the periods are $\ell= 1$  (solution 1). Among solutions 2, 3
and 4, the worst one is solution 3.  Solution 2 has the lower $\chi^2$
value, but 3 of the 4 periods  observed in PG 1351+489 are fitted with
theoretical periods associated to  $\ell= 2$ modes, and it is true in
particular for the period at $\sim 489$ s  which has the largest
amplitude. In view of geometrical  cancellation effects at the stellar
surface \cite{1977AcA....27..203D},  it is improbable that the
largest-amplitude mode with period at $\sim 489$ s corresponds to
$\ell= 2$.  Solution 4  has a quality comparable than  solution 2, but
in this case the period of the highest  amplitude mode is reproduced
by a theoretical $\ell= 1$ period. In addition,  the effective
temperature of this solution is in agreement with  the spectroscopic
derivation for $T_{\rm eff}$. For these reasons, we adopt solution 4
as the asteroseismological model for PG 1351+489. 

In Table \ref{table2} we show the characteristics of our best-fitting
model and the parameters derived from spectroscopy. Note that the
effective  temperature of the model is in good agreement with that of
PG 1351+489 inferred by  spectroscopy, but the surface gravity  is
substantially larger than the spectroscopic  one. We emphasize that
the quoted errors in the parameters of the  asteroseismological
models are just \emph{internal} errors. The assessment of the
\emph{true} uncertainties is beyond the scope of the present paper,
but we envisage that they could be  somewhat larger than those
included in Table \ref{table2}.

\begin{table}
\centering
\caption{Characteristics of PG 1351+489 derived from the spectroscopic
  analysis \cite{2011ApJ...737...28B}  and  results  of our  best
  asteroseismological   model.  $X_{\rm C}$ and $X_{\rm O}$ are the
  fractional abundances of C and O, respectively, at the stellar
  center. The  quoted uncertainties  in  the asteroseismological
  model  are the  \emph{internal} errors of our period-fitting
  procedure.}
\begin{tabular}{lcc}
\hline
\hline
\noalign{\smallskip}
 Quantity                        & Spectroscopy      & Asteroseismology           \\  
\noalign{\smallskip}
\hline
\noalign{\smallskip}
$T_{\rm eff}$ [K]                & $26\,010 \pm 1536$  & $25\, 775 \pm 150$   \\
$M_*/M_{\odot}$                   & $0.558 \pm 0.027$   & $0.664 \pm 0.013$    \\
$\log g$                        & $7.91 \pm 0.07$     & $8.103 \pm 0.020$     \\
$\log(R_*/R_{\odot})$             &    ---              & $-1.912 \pm 0.015$   \\
$\log(L_*/L_{\odot})$             &    ---              & $-1.244 \pm 0.03$     \\
$M_{\rm He}/M_*$                 &    ---              & $3.63 \times 10^{-3}$  \\
$X_{\rm C},X_{\rm O}$ (center)    &    ---              & $0.32, 0.65$ \\
\noalign{\smallskip}
\hline
\hline
\end{tabular}
\label{table2}
\end{table}

\begin{table}
\centering
\caption{The observed periods and rates of period change of PG 1351+489 
(columns 1 and 5), and the 
theoretical periods and rates of period changes of the asteroseismological 
model (columns 2 and 6), 
along with the corresponding  $(\ell, k)$ mode  identification (columns 3 and 4).}
\begin{tabular}{cccccc}
\hline
\hline
\noalign{\smallskip}
$\Pi^{\rm o}$        & $\Pi^{\rm t}$       & $\ell$ & $k$ & $\dot{\Pi}^{\rm o}$  & $\dot{\Pi}^{\rm t}$\\
\noalign{\smallskip}
$[$s$]$              &  $[$s$]$            &        &    & $[10^{-13} $s/s$]$   & $[10^{-13}  $s/s$]$\\  
\noalign{\smallskip}
\hline
\noalign{\smallskip}
335.26    & 336.81   & 2 &  13  & ---            & 0.60 \\
489.33    & 489.47   & 1 &  11  & $2.0 \pm 0.9$  & 0.81 \\
584.68    & 586.99   & 2 &  25  & ---            & 1.02 \\
639.63    & 639.37   & 1 &  15  & ---            & 1.19 \\
\noalign{\smallskip}
\hline
\hline
\end{tabular}
\label{table3}
\end{table}

A detailed  comparison of the observed  periods  in PG 1351+489 with
the theoretical periods of the best-fit asteroseismological model is
provided  in  Table  \ref{table3}.  We include also the observed and
theoretical  rates of period changes for each mode. The model very closely
reproduces the observed periods. Interestingly, the observed rate of
change of the $\sim 489.33$ s  period is more than two times larger
than the theoretically expected value.  If we assume that the rate of
period change of this mode reflects the evolutionary time-scale of the
star, and that the asteroseismological model truly represents the real
star, then the disagreement between the observed and theoretical
values of $\dot{\Pi}$ would  indicate that PG 1351+489 could be cooling
faster than that predicted by the standard theory of white dwarf
evolution. 

\begin{figure}
\centering
\includegraphics[clip, width=0.7\textwidth, angle=0]{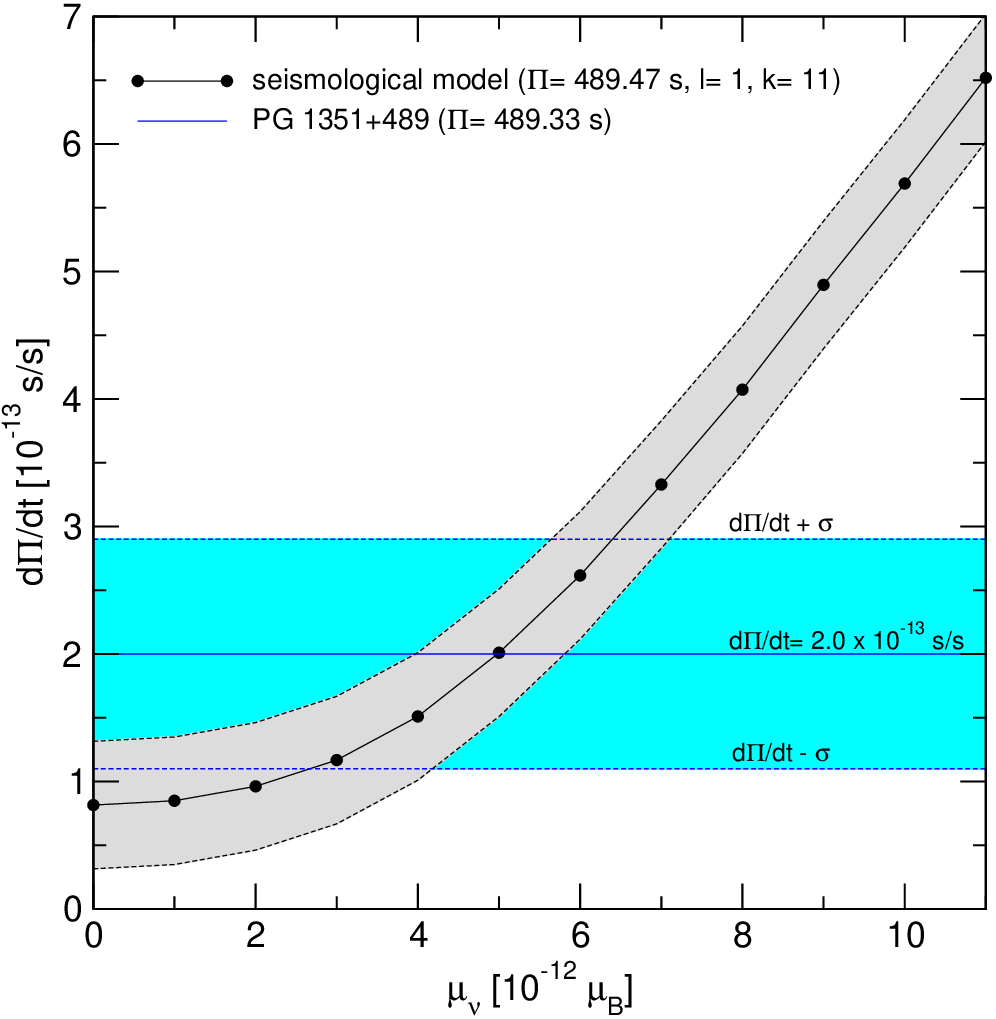}
\caption{The rate of period change of the $\ell= 1$ and $k= 11$  mode in terms
of the neutrino magnetic dipole moment, corresponding to 
the asteroseismological model for PG 1351+489 
($M_*= 0.664 M_{\odot}$ and $T_{\rm eff}= 25\,775$ K).
The estimate of the rate of period 
change of the 489 s period \cite{2011MNRAS.415.1220R} of PG 1351+489 and its 
uncertainties are shown with thin blue horizontal lines.}
\label{dpdt-mu-0664}
\end{figure}

Having found that the theoretically expected rate of change of period
with time for the $k$ = 11 mode is somewhat smaller than the value
measured for PG 1351+489, suggesting the existence of some additional
cooling mechanism in this star, we now assume that this additional
cooling is entirely attributable to the existence of a neutrino
magnetic dipole moment\footnote{We neglect the emission of axions or
  other weakly interacting particles.}.  In this way, we can establish
an upper bound for the possible value of $\mu_\nu$.  Specifically, we
considered different values of $\mu_\nu/\left(10^{12} \mu_{\rm
  B}\right)$  in the range $(0-11)$ and assumed the same structural
parameters $(T_{\rm eff}, M_*)$ as the asteroseismological  model.
This is a valid procedure because, as shown in Sect. \ref{effects},
the pulsation periods do not depend on the magnitude of $\mu_\nu$,
which means that the asteroseismological model for PG 1351+489 in  the
case of a non-zero neutrino magnetic dipole moment is still valid.  In
Fig. \ref{dpdt-mu-0664} we display  the theoretical value  of
$\dot{\Pi}$ corresponding to  the period $\Pi\sim  489$~s for
increasing  values of the neutrino magnetic dipole moment (black
solid curve and dots).  The  dashed curves   embracing the solid curve
represent   the  uncertainty   in  the   theoretical   value  of
$\dot{\Pi}$, $\varepsilon_{\dot{\Pi}}= 0.5 \times 10^{-13}$~s/s. This
value has been roughly estimated for the case in which $\mu_\nu= 0$
by comparing the $\dot{\Pi}$ of the $\sim 489$ s mode of the
asteroseismological  model with the value of $\dot{\Pi}$ of the
eigenmodes with similar periods  belonging to the  sequences with
$M_*= 0.609 M_{\odot}$  and $M_*= 0.741 M_{\odot}$ at  similar
effective temperature ($T_{\rm eff} \sim 25\,800$ K).  Additionally,
we  assume that the  uncertainties  do  not depend  on  the  value of
$\mu_\nu$.    The horizontal  (blue) solid  line  indicates  the
observed  value  of $2.0\times   10^{-13}$~s/s,   whilst   its
corresponding   $1\sigma$ uncertainties ($\pm 0.9 \times
10^{-13}$~s/s) \cite{2011MNRAS.415.1220R}   are shown  with  dashed
lines.  If  one standard  deviation from  the  observational value  is
considered,  we conclude  that the  neutrino magnetic dipole moment
is $\mu_\nu \lesssim 7 \times 10^{-12} \mu_{\rm B}$.  This  upper
limit ​is  compatible  with  those obtained in the previous section
without considering an  asteroseismological model ($\mu_\nu \lesssim 9
\times 10^{-12} \mu_{\rm B}$).

\section{Summary and conclusions}
\label{conclusions}

In this paper we have derived for the first time an upper bound  to
the neutrino magnetic dipole moment employing the DBV white dwarf
pulsator PG 1351+489, for which an estimate of the rate of period
change of the dominant pulsation  period is available.  Our analysis
is based on state-of-the-art evolutionary DB white dwarf models
consistent with the history of progenitor stars.   
We obtain a bound of $\mu_\nu \lesssim 9 \times 10^{-12} \mu_{\rm B}$ 
when we rely  on
the spectroscopic values of $T_{\rm eff}$ and $\log g$ of this star,
and a limit of $\mu_\nu \lesssim 7 \times 10^{-12} \mu_{\rm B}$ if we
consider an asteroseismological model that closely reproduces   its
pulsation periods. We adopt a conservative limit of 
$\mu_\nu \lesssim 10^{-11} \mu_{\rm B}$ as the main result of 
our analysis, which reinforces the more 
restrictive upper bound derived on the basis of the 
WDLF \cite{2014A&A...562A.123M},
$\mu_\nu < 5 \times 10^{-12} \mu_{\rm B}$,  and that derived from an
analysis of red giants from the color-magnitude  diagram (CMD) of the
Galactic globular cluster M5,   $\mu_\nu < 4.5 \times 10^{-12}
\mu_{\rm B}$ (95 \% CL)
\cite{2013A&A...558A..12V,2013PhRvL.111w1301V}. 
 Notwithstanding the rate of
change of the dominant period in PG 1351+489 we are using in our analysis 
is still not a definitive measurement, the derived constraint for
$\mu_\nu$ does not depend on actually measuring the rate of period change. 
This is because our constraint is based on a measured upper limit to 
the period drift, not on the period drift itself.
Of course, if the uncertainty on the measured rate of period change 
would improve, this would also improve the limits on $\mu_\nu$, or could 
even lead to a clear requirement for additional cooling effects.

\acknowledgments  
We thank the anonymous referee for his/her
encouraging report. Part  of this work  was supported by  AGENCIA
through the Programa  de Modernizaci\'on Tecnol\'ogica BID
1728/OC-AR,  and by the PIP   112-200801-00940   grant    from
CONICET.    EG-B acknowledges support from the AGAUR, from MCINN
(Grant AYA2011-23102), and from the European Union FEDER fund. This
research has made use of NASA's Astrophysics Data System.

\bibliographystyle{JHEP}
\bibliography{paper-mmagnetico.bib}

\end{document}